\documentclass[12pt]{article}
\usepackage[colorlinks=true,urlcolor=blue,anchorcolor=blue,%
citecolor=blue,filecolor=blue,linkcolor=blue,menucolor=blue,% 
linktocpage=true,pdfproducer=medialab]{hyperref}
\usepackage{jhep-mod-2018}
\usepackage{graphics}
\usepackage{tikz}
\newcommand{\orcidicon}{%
	\begin{tikzpicture}
	\draw[lime, fill=lime] (0,0) 
		circle [radius=0.16] 
		node[white] {{\fontfamily{qag}\selectfont \tiny ID}};
	\draw[white, fill=white] (-0.0625,0.095) 
		circle [radius=0.007];
	\end{tikzpicture}
	\hspace{-3mm}
}
\newcommand\orcidMatt{{\href{https://orcid.org/0000-0003-1088-6485}{\orcidicon}}}
\begin{document}
%========================================================%----------------------------------------------------------------------------
\title{\huge The exponential metric represents\\ a traversable wormhole}
%----------------------------------------------------------------------------
\author{\Large Petarpa Boonserm$\,^{1,2}$, Tritos Ngampitipan$\,^{3}$, \\
Alex Simpson$\,^{4}$, {\sf  and} Matt Visser$\,^{4}$\orcidMatt}
%========================================================%========================================================
%========================================================%========================================================
\affiliation{
$^{1}$  Department of Mathematics and Computer Science, Faculty of Science,\\
\null\qquad Chulalongkorn University,  Bangkok 10330, Thailand.  
}
\affiliation{
$^2$ Thailand Center of Excellence in Physics, Ministry of Education, \\
\null\qquad Bangkok 10400, Thailand.
}
\affiliation{
$^{3}$  Faculty of Science, Chandrakasem Rajabhat University,
 Bangkok 10900, Thailand.
}
\affiliation{$^{4}$ School of Mathematics and Statistics, Victoria University of Wellington, \\
\null\qquad PO Box 600, Wellington 6140, New Zealand}
%========================================================%========================================================
\emailAdd{petarpa.boonserm@gmail.com}
\emailAdd{tritos.ngampitipan@gmail.com}
\emailAdd{alex.simpson@sms.vuw.ac.nz}
\emailAdd{matt.visser@sms.vuw.ac.nz}
%========================================================%========================================================

\abstract{

\parindent0pt
\parskip7pt

For various reasons a number of authors have mooted an ``exponential form'' for the spacetime metric:
\[
ds^2 = - e^{-2m/r} dt^2 + e^{+2m/r}\{dr^2 + r^2(d\theta^2+\sin^2\theta \, d\phi^2)\}.
\]
While the weak-field behaviour matches nicely with weak-field general relativity,
and so also automatically matches nicely with the Newtonian gravity limit, the strong-field behaviour is markedly different.  Proponents of these exponential metrics have very much focussed on the absence of horizons --- it is certainly clear that this geometry does not represent a black hole. However, the proponents of these exponential metrics have failed to note that instead one is dealing with a traversable wormhole --- with all of the interesting  and potentially problematic features that such an observation raises.
If one wishes to replace all the black hole candidates astronomers have identified with traversable wormholes, then certainly a careful phenomenological analysis of this quite radical proposal should be carried out.

\bigskip
{\sc Date:} 10 May 2018; 17 May 2018; 21 May 2018; \LaTeX-ed \today


\noindent\textbf{\sc Pacs}: 04.20.-q; 04.20.-q; 04.20.Jb; 04.70.Bw
}

%----------------------------------------------------------------------------
\maketitle
%----------------------------------------------------------------------------
%----------------------------------------------------------------------------
\def\d{{\mathrm{d}}}
\def\V{{\mathcal V}}
%----------------------------------------------------------------------------
%----------------------------------------------------------------------------
%----------------------------------------------------------------------------
%----------------------------------------------------------------------------
\clearpage
%----------------------------------------------------------------------------
%----------------------------------------------------------------------------
\section{Introduction}
%========================================================
%---------------------------------------------------------------------------------------------------------------------------------------------
\label{S:intro}
%---------------------------------------------------------------------------------------------------------------------------------------------

The so-called ``exponential metric''
\begin{equation}
ds^2 = - e^{-2m/r} dt^2 + e^{+2m/r}\{dr^2 + r^2(d\theta^2+\sin^2\theta \, d\phi^2)\},
\end{equation}
has now been in circulation for some 60 years~\cite{Yilmaz:1958, Yilmaz:1971, Yilmaz:1973, Clapp:1973, Rastall:1975, Fennelly:1976, Misner:1995, Alley:1995, MECO:1999, Robertson:1999, Ibison:2006a, Ibison:2006b, BenAmots:2007, Svidzinsky:2009, Martinis:2010, BenAmots:2011, Svidzinsky:2015, Aldama:2015, MECO:2016};
at least since 1958.
Motivations for considering this metric vary quite markedly, (even between different papers written by the same author), and the theoretical ``justifications''  advanced for considering this particular space-time metric are often rather dubious.  Nevertheless a small segment of the community has consistently advocated for this particular spacetime metric for over 60 years, with significant activity continuing up to the present day.
Regardless of one's views regarding the purported theoretical ``justifications''  for this metric, one can simply take this metric as given, and then try to understand its phenomenological properties; some of which are significantly problematic.

A particularly attractive feature of this exponential metric is that in weak fields, ($2m/r\ll 1$), one has
\begin{equation}
ds^2 = \{- dt^2 +dr^2 + r^2(d\theta^2+\sin^2\theta \, d\phi^2)\} +
{2m\over r} \{dt^2 +dr^2 + r^2(d\theta^2+\sin^2\theta \, d\phi^2)\}.
\end{equation}
That is
\begin{equation}
g_{ab} = \eta_{ab} + {2m\over r} \delta_{ab}.
\end{equation}
This exactly matches the lowest-order weak-field expansion of general relativity, and so this exponential metric will automatically pass all of the standard lowest-order weak-field tests of general relativity. However strong-field behaviour,  ($2m/r\gg 1$),  and even medium-field behaviour,  ($2m/r \sim 1$), is markedly different.

The exponential metric has no horizons, $g_{tt}\neq0$, and so is not a black hole. On the other hand, it does not seem to have been previously remarked that the exponential metric describes a traversable wormhole in the sense of Morris and Thorne~\cite{Morris-Thorne,MTY,Visser:1989a,Visser:1989b,Visser:book, Cramer:1994, Poisson:1995, Hochberg:1997, Visser:1997, Hochberg:1998, Hochberg:1998b, Barcelo:1999, Barcelo:2000, Dadhich:2001, Visser:2003,Lemos:2003,Kar:2004,Lobo:2005,Sushkov:2005,Garcia:2011, Bhawal, Arias, Gao, Maldacena, Willenborg, Sahoo}. We shall demonstrate that the exponential metric has a wormhole throat at $r=m$, with the region $r<m$ corresponding to an infinite-volume ``other universe'' that exhibits the ``underhill effect''; time runs slower on the other side of the wormhole throat.

%\clearpage

%------------------------------------------------
\section{Traversable wormhole throat}\label{S:throat}
%------------------------------------------------

Consider the area of the spherical surfaces of constant $r$ coordinate:
\begin{equation}
A(r) = 4\pi r^2 e^{2m/r}.
\end{equation}
Then
\begin{equation}
{dA(r)\over dr} = 8\pi (r-m) e^{2m/r};
\end{equation}
and
\begin{equation}
{d^2A(r)\over dr^2} = 8\pi  e^{2m/r} \left(1-{2m\over r} +{2m^2\over r^2}\right)
=
8\pi  e^{2m/r} \left\{\left(1-{m\over r}\right)^2  +{m^2\over r^2}\right\} > 0.
\end{equation}
That is: The area is a concave function of the $r$ coordinate, and has a minimum at $r=m$, where it satisfies the ``flare out'' condition  $A''|_{r=m} = +8\pi e^2 > 0$.
Furthermore, all metric components are finite at $r=m$, and the diagonal components are non-zero. This is sufficient to guarantee that the surface $r=m$ is a traversable wormhole throat,  in the sense of Morris and Thorne~\cite{Morris-Thorne,MTY,Visser:1989a,Visser:1989b,Visser:book, Cramer:1994, Poisson:1995, Hochberg:1997, Visser:1997, Hochberg:1998, Hochberg:1998b, Barcelo:1999, Barcelo:2000, Dadhich:2001, Visser:2003,Lemos:2003,Kar:2004,Lobo:2005,Sushkov:2005,Garcia:2011, Bhawal, Arias, Gao, Maldacena, Willenborg, Sahoo}. There is a rich phenomenology of traversable wormhole physics that has been developed over the last 30 years, (since the Morris-Thorne paper~\cite{Morris-Thorne}), much of which can be readily adapted (\emph{mutatis mutandi}) to the exponential metric.

%------------------------------------------------
\section{Comparison: Exponential versus Schwarzschild}\label{S:compare}
%------------------------------------------------

Let us briefly compare the exponential and Schwarzschild metrics. 

%------------------------------------------------
\subsection{Isotropic coordinates}\label{S:isotropic}
%------------------------------------------------

In isotropic coordinates the Schwarzschild spacetime is 
\begin{equation}
ds_\mathrm{Sch}^2 = - \left(1-{m\over2r}\over1+{m\over2r}\right)^2 dt^2
+ \left(1+{m\over2r}\right)^4 \{dr^2 +  r^2(d\theta^2+\sin^2\theta \, d\phi^2)\},
\end{equation}
which we should compare with the exponential metric in isotropic coordinates
\begin{equation}
ds^2 = - e^{-2m/r} dt^2 + e^{+2m/r}\{dr^2 + r^2(d\theta^2+\sin^2\theta \, d\phi^2)\}.
\end{equation}
It is clear that in the Schwarzschild spacetime there is a horizon present at $r=\frac{m}{2}$.
Recalling that the domain for the $r$-coordinate in the isotropic coordinate system for Schwarzschild is $r\in (0, +\infty)$, we see that the horizon also corresponds to where the area of spherical constant-$r$ surfaces is minimised:
\begin{align}
A(r) & =4\pi r^2\left(1+\frac{m}{2r}\right)^{4}; \\
\\
\frac{dA(r)}{dr} & =8\pi r\left(1-\frac{m}{2r}\right)\left(1+\frac{m}{2r}\right)^{3}; \\
\\
\frac{d^{2}A(r)}{dr^2} & =
8\pi\left(1+\frac{m}{2r}\right)^{2}\left(\frac{3}{4}\left(\frac{m}{r}\right)^{2}-\frac{m}{r}+1\right) \ .
\end{align}
So for the Schwarzschild geometry in isotropic coordinates the area has a minimum at $r=\frac{m}{2}$, where $A'\vert_{r=\frac{m}{2}}=0$, and $A''\vert_{r=\frac{m}{2}}=64\pi >0$.
While this satisfies the ``flare-out'' condition the corresponding wormhole (the Einstein--Rosen bridge) is \emph{non-traversable} due to the presence of the horizon. 

In contrast the geometry described by the exponential metric clearly has no horizons, since $\forall \ r\in (0,+\infty)$ we have $\exp\left({\frac{-2m}{r}}\right)\neq 0$. As already demonstrated, there is a  traversable wormhole throat located at $r=m$, where the area of the spherical surfaces is minimised, and the ``flare out'' condition is satisfied, in the \emph{absence} of a horizon. Thus the Schwarzschild horizon at $r=\frac{m}{2}$ in isotropic coordinates is replaced by a wormhole throat at $r=m$ in the exponential metric.

Furthermore, for the exponential metric, since $\exp\left({\frac{-2m}{r}}\right) >0$ is monotone decreasing as $r\to0$, it follows that proper time evolves increasingly slowly as a function of coordinate time as one moves closer to the centre $r\to 0$.  

%------------------------------------------------
\subsection{Curvature coordinates}\label{S:curvature}
%------------------------------------------------

To go to so-called ``curvature coordinates",  (often called ``Schwarzschild curvature coordinates''), for the exponential metric we make the coordinate transformation
\begin{equation}
r_s = r \, e^{m/r}; \qquad\qquad dr_s = e^{m/r}  \, (1 - m/r) dr.
\end{equation}
So for the exponential metric in curvature coordinates
\begin{equation}
ds^2 = - e^{-2m/r} dt^2 + {dr_s^2\over (1-m/r)^2} + r_s^2(d\theta^2+\sin^2\theta \, d\phi^2).
\end{equation}
Here $r$ is regarded as an implicit function of $r_s$.
Note that as the isotropic coordinate $r$  ranges over the interval $(0,\infty)$, the curvature coordinate $r_s$ has a minimum at $r_s=m\,e$. In fact for the exponential metric the curvature coordinate $r_s$ double-covers the interval $r_s\in[m\, e,\infty)$, first descending from $\infty$ to $m\, e$ and then increasing again to $\infty$.
Indeed, looking for the minimum of the coordinate $r_{s}$:
\begin{align}
\frac{dr_{s}}{dr}  =e^{\frac{m}{r}}\left(1-\frac{m}{r}\right)
\qquad\Longrightarrow \qquad 
\left.\frac{dr_{s}}{dr}\right\vert_{r=m}=0.
\end{align}
So we have a stationary point at $r=m$, which corresponds to $r_{s}=m\,e$, and furthermore 
\begin{align}
\frac{d^{2}r_{s}}{dr^{2}}  =\frac{m^2}{r^{3}}\;e^{\frac{m}{r}} 
\qquad \Longrightarrow \qquad  \left.\frac{d^{2}r_{s}}{dr^{2}}\right\vert_{r=m}>0 \ .
\end{align}
The curvature coordinate $r_{s}$ therefore has a minimum at $r_{s}=m\,e$, and in these curvature  coordinates the exponential metric exhibits a wormhole throat at $r_{s}=m\,e$.

Compare this with the Schwarzschild metric in curvature coordinates:
\begin{equation}
ds_\mathrm{Sch}^2 = - (1-2m/r_s) dt^2 + {dr_s^2\over 1-2m/r_s} + r_s^2(d\theta^2+\sin^2\theta \, d\phi^2).
\end{equation}
By inspection it is clear that there is a horizon at $r_{s}=2m$, since at that location  $g_{tt}\vert_{r_{s}=2m}=0$. For the Schwarzschild metric the isotropic and curvature coordinates are related by $r_s = r \left(1+{m\over2r}\right)^2$. 

\enlargethispage{40pt}
If for the exponential metric one really wants the fully explicit inversion of $r$ as a function of $r_s$, then observe
\begin{equation}
r =r_s \exp\left( W(-m/r_s)\right) = - {m\over W(-m/r_s)}.
\end{equation}
Here $W(x)$ is ``appropriate branch'' of the Lambert $W$ function --- implicitly defined by the relation $W(x)\, e^{W(x)}=x$.
This function has a convoluted 250-year history; only recently has it become common to view it as one of the standard ``special functions''~\cite{Corless}. Applications vary~\cite{Corless,Valluri:00}, including combinatorics (enumeration of rooted trees)~\cite{Corless},  delay differential equations~\cite{Corless}, falling objects subject to linear drag~\cite{Vial:12}, evaluating the numerical constant in Wien's displacement law~\cite{Stewart:11, Stewart:12}, quantum statistics~\cite{Valluri:09},
the distribution of prime numbers~\cite{primes}, constructing the ``tortoise'' coordinate for Schwarzschild black holes~\cite{tortoise}, \emph{etcetera}.

In terms of the $W$ function and the curvature coordinate $r_s$ the explicit version of the exponential metric becomes
\begin{equation}
ds^2 = - e^{2W(-m/r_s)} dt^2 + {dr_s^2\over (1+W(-m/r_s))^2} + r_s^2(d\theta^2+\sin^2\theta \, d\phi^2).
\end{equation}
The $W_0(x)$ branch corresponds to the region $r>m$ outside the wormhole throat;
whereas the $W_{-1}(x)$ branch corresponds to the region $r<m$ inside the wormhole.
The Taylor series for $W_0(x)$ for $|x| < e^{-1}$ is~\cite{Corless}
\begin{equation}
W_0(x) = \sum_{n=1}^\infty {(-n)^{n-1} x^n\over n!}.
\end{equation}
A key asymptotic formula for $W_{-1}(x)$ is~\cite{Corless}.
\begin{equation}
W_{-1}(x) =  \ln(-x) - \ln(-\ln(-x)) + o(1);  \qquad (x\to 0^-).
\end{equation}
The two real branches meet at $W_0(-1/e)=W_{-1}(-1/e)=-1$, and in the vicinity of that meeting point
\begin{equation}
W(x) =  -1 + \sqrt{2(1+ex)} - {2\over3} (1+ex) + O[(1+ex)^{3/2}].
\end{equation}
More details regarding the Lambert $W$ function can be found in Corless~\emph{et al}, see~\cite{Corless}.

%------------------------------------------------
\section{Curvature tensor}\label{S:Pi}
%------------------------------------------------

The curvature components for the exponential metric (in isotropic coordinates) are easily computed.
For the Riemann tensor the non-vanishing components are:
\begin{eqnarray}
R^{tr}{}_{tr}&=&-2R^{t\theta}{}_{t\theta} = -2 R^{t\phi}{}_{t\phi} = {2m(r-m)e^{-2m/r}\over r^4};\\
R^{r\theta}{}_{r\theta} &=& R^{r\phi}{}_{r\phi} = -{m e^{-2m/r}\over r^3};\\
R^{\theta\phi}{}_{\theta\phi} &=&  {m(2r-m)e^{-2m/r}\over r^4}.\\
\end{eqnarray}
For the Weyl tensor the non-vanishing components are even simpler:
\begin{equation}
C^{tr}{}_{tr}=-2C^{t\theta}{}_{t\theta} = -2 C^{t\phi}{}_{t\phi}
=-2C^{r\theta}{}_{r\theta} = -2 C^{r\phi}{}_{r\phi} =
C^{\theta\phi}{}_{\theta\phi} = {2\over3} {m(3r-2m)e^{-2m/r}\over r^4}.\\
\end{equation}

For the Ricci and Einstein tensors:
\begin{eqnarray}
R^a{}_{b} &=& -{2m^2 e^{-2m/r}\over r^4} \; \text{diag}\{0,1,0,0\}^a{}_{b};
\\
R &=& -{2m^2 e^{-2m/r}\over r^4};
\\
G^a{}_b &=& {m^2 e^{-2m/r}\over r^4} \; \text{diag}\{1,-1,1,1\}^a{}_{b}.
\end{eqnarray}
For the Kretschmann and other related scalars we have
\begin{equation}
R_{abcd}\, R^{abcd} = {4m^2(12r^2-16mr+7m^2)e^{-4m/r}\over r^8};
\end{equation}
\begin{equation}
C_{abcd}\, C^{abcd} = {16\over3} {m^2(3r-2m)^2 e^{-4m/r}\over r^8};
\end{equation}
\begin{equation}
R_{ab}\, R^{ab} = R^2 = {4m^4 e^{-4m/r}\over r^8}.
\end{equation}
All of the curvature components and scalar invariants exhibited above are finite everywhere in the exponential spacetime --- in particular they are finite at the throat ($r=m$) and decay to zero both as $r\to\infty$ and as $r\to0$. They take on maximal values near the throat, at $r=\hbox{(dimensionless number)}\times m$.

%-------------------------------------------------
\section{Ricci convergence conditions}
%--------------------------------------------------

Since most of the advocates of the exponential metric are typically not working within the framework of general relativity, and typically do not want to enforce the Einstein equations, the standard energy conditions are to some extent moot. In the usual framework of general relativity the standard energy conditions are useful because they feed back into the Raychaudhuri equations and its generalizations, and so give information about the focussing and defocussing of geodesic congruences. In the absence of the Einstein equations one can instead impose conditions directly on the Ricci tensor.

Specifically, a Lorentzian spacetime is said to satisfy the timelike, null, or spacelike Ricci convergence condition if for all timelike, null, or spacelike vectors $t^{a}$ one has:
\begin{equation}
R_{ab}\; t^{a}t^{b}\geq 0 \ .
\end{equation}
For the exponential metric one has
\begin{equation}
R_{ab} = -{2m^2 \over r^4} \; \text{diag}\{0,1,0,0\}_{ab}.
\end{equation}
So the Ricci convergence condition  amounts to
\begin{equation}
R_{ab}\; t^{a}t^{b} =  -{2m^2 \over r^4} (t^r)^2 \leq 0.
\end{equation}
This clearly will not be satisfied for all timelike, null, or spacelike vectors $t^{a}$.
Specifically, the  violation of the null Ricci convergence condition is crucial for understanding the flare out at the throat of the traversable wormhole~\cite{Visser:book}. 

%------------------------------------------------
\section{Effective refractive index --- lensing properties}\label{S:refractive}
%------------------------------------------------
\enlargethispage{40pt}
The exponential metric can be written in the form
\begin{equation}
ds^2 = e^{2m/r} \left\{ - e^{-4m/r} dt^2 + \{dr^2 + r^2(d\theta^2+\sin^2\theta \, d\phi^2)\} \right\}.
\end{equation}
If we are only interested in photon propagation, then the overall conformal factor is irrelevant, and we might as well work with
\begin{equation}
d\hat s^2 =- e^{-4m/r} dt^2 + \{dr^2 + r^2(d\theta^2+\sin^2\theta \, d\phi^2)\}.
\end{equation}
That is
\begin{equation}
d\hat s^2 =- e^{-4m/r} dt^2 + \{dx^2 +dy^2 +dz^2\}.
\end{equation}
But this metric has a very simple physical interpretation: It corresponds to a coordinate speed of light $c(r) = e^{-2m/r}$, or equivalently an effective refractive index
\begin{equation}
n(r) = e^{2m/r}.
\end{equation}
This effective refractive index is well defined all the way down to $r=0$, and (via Fermat's principle of least time) completely characterizes the focussing/defocussing of null geodesics. 
This notion of ``effective refractive index'' for the gravitational field has in the weak field limit been considered in~\cite{refractive}, and in the strong-field limit falls naturally into the ``analogue spacetime'' programme~\cite{LRR,LNP}.

Compare the above with Schwarzschild spacetime in isotropic coordinates where the effective refractive index is
\begin{equation}
n(r) =  {(1+{m\over2r})^3\over|1-{m\over2r}|}.
\end{equation}
The two effective refractive indices have the same large-$r$ limit, $n(r)\approx 1 + {2m\over r}$, but differ markedly once $r\lesssim m/2$. 

%---------------------------------------------------------------------
\begin{figure}[!htb]
\begin{center}
\includegraphics[scale=0.35]{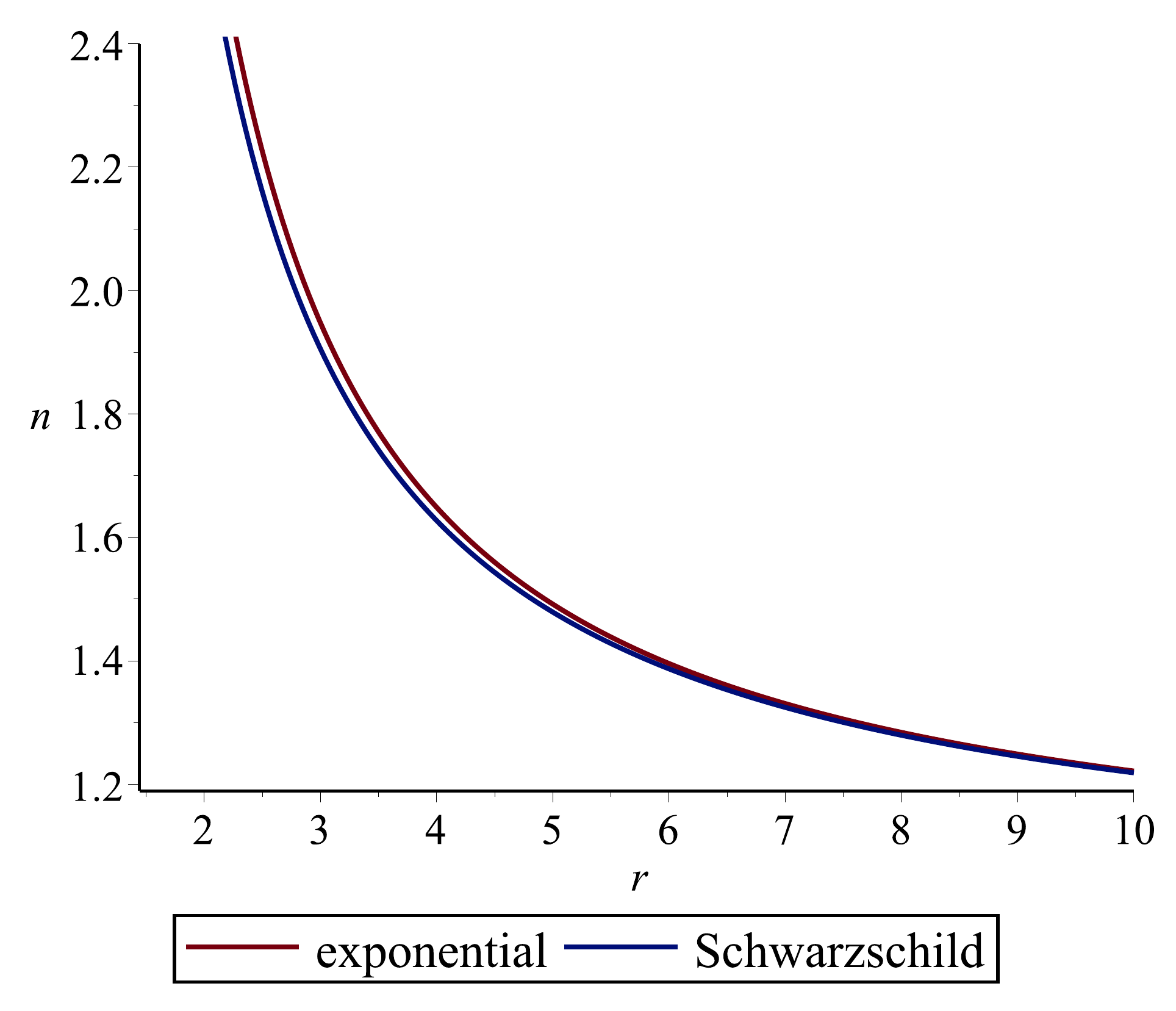}\qquad
\includegraphics[scale=0.35]{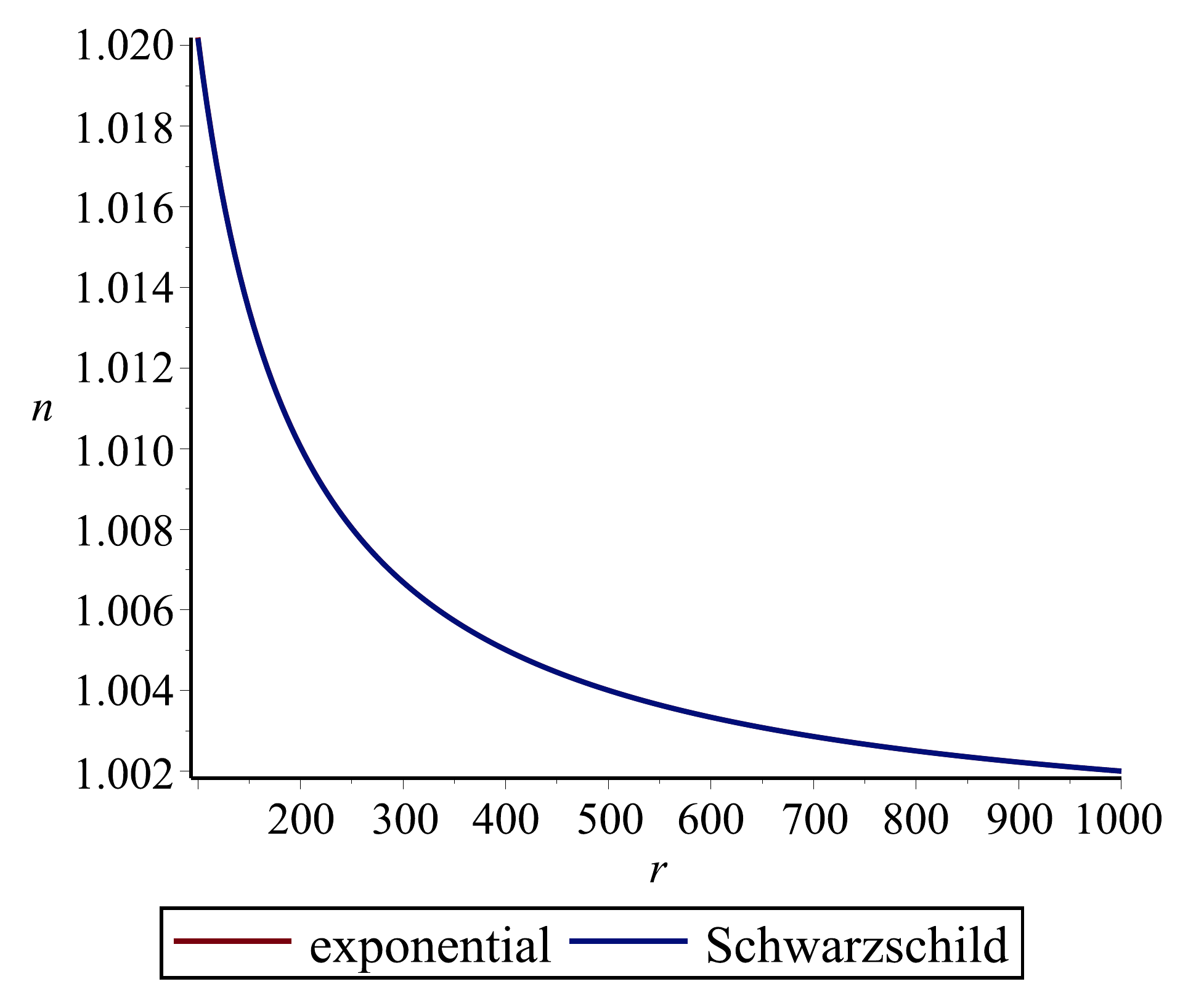}
\includegraphics[scale=0.45]{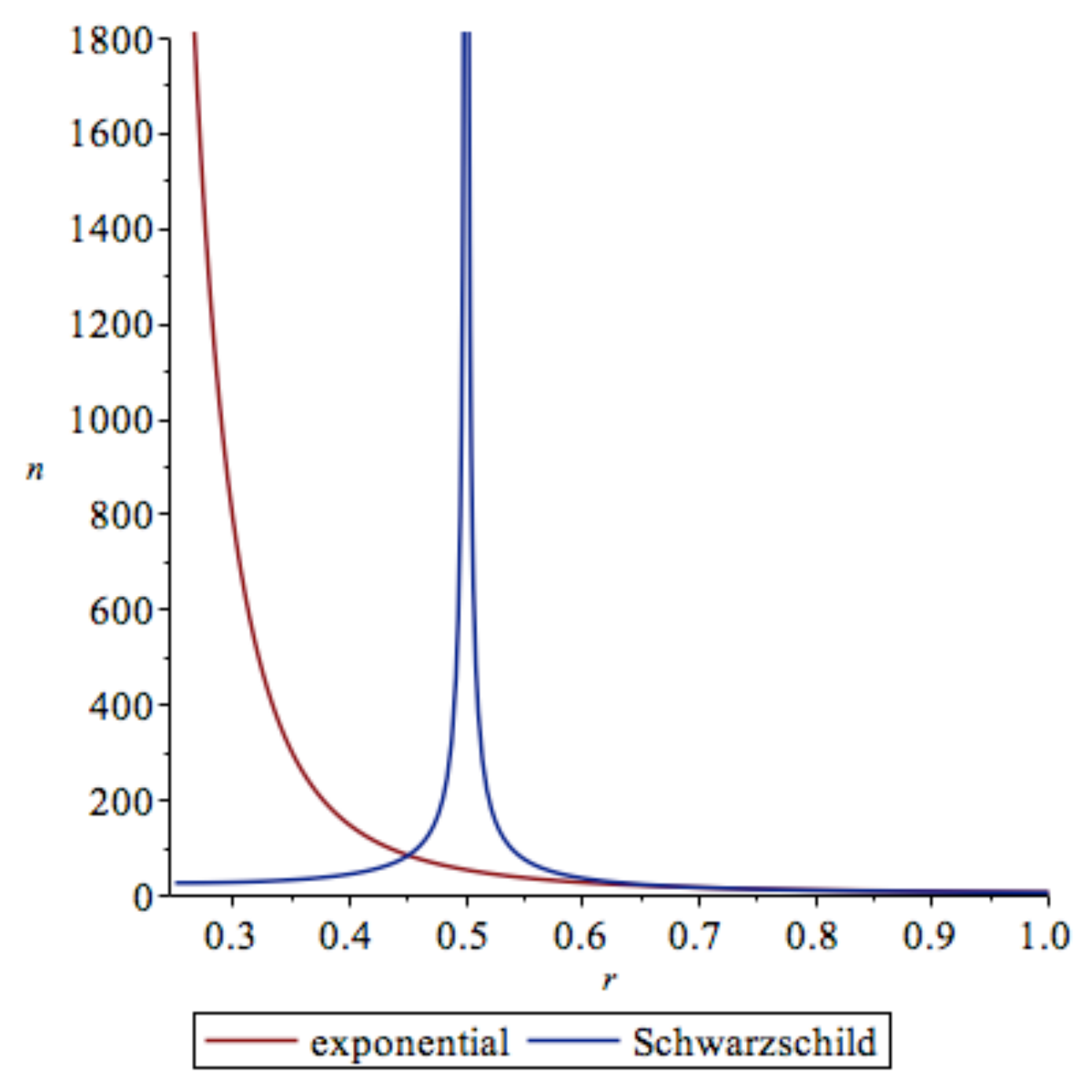}
\end{center}
{\caption{{The graph shows the refractive index for the exponential metric compared to the Schwarzschild metric in the isotropic coordinate. The parameter $m = 1$. The left panel is for relatively small $r\gtrsim2m$ and the right panel is for large $r$. The bottom panel is for the strong field region $r\sim m/2$.}}\label{pott}}
\end{figure}
%--------------------------------------------------------------------

From the graphs presented in Figure \ref{pott}, we can see that the refractive index for the exponential metric is greater than that of the Schwarzschild metric in the isotropic coordinate at tolerably small $r\gtrsim 2m$. For large $r$, they converge to each other and hence are asymptotically equal. In the strong field region they differ radically. Observationally, once you get close enough to where you would have expected to see the Schwarzschild horizon, the lensing properties differ markedly. 

%------------------------------------------------
\section{ISCO (innermost stable circular orbit) and photon sphere}
%------------------------------------------------

For massive particles, it is relatively easy to find the innermost stable circular orbit (ISCO) for the exponential metric;  while for massless particles such as photons there  is a unique unstable circular orbit. These can then be compared with Schwarzschild spacetime.
We emphasize that the notion of ISCO depends only on the geodesic equations, not on the assumed field equations chosen for setting up the spacetime. 
Since Schwarzschild ISCOs for massive particles at $r_s=6m$ have already been seen by astronomers;
this might place interesting bounds somewhat restraining the exponential-metric enthusiasts.
Additionally, the Schwarzschild unstable circular photon orbit for massless particles is at $r_s=3m$ (the photon sphere); the equivalent for the exponential metric is relatively easy to find.

\def\L{{\mathcal{L}}}
To determine the circular orbits, consider the affinely parameterized tangent vector to the worldline of a massive or massless particle
\begin{equation}
g_{ab} {dx^a\over d\lambda} {d x^b\over d\lambda}
= -e^{-2m/r} \left(dt \over d\lambda\right)^2
+ e^{2m/r} \left\{ \left(dr \over d\lambda\right)^2 + r^2\left[\left(d\theta \over d\lambda\right)^2+ \sin^2\theta\left(d\phi \over d\lambda\right)^2\right] \right\} = \epsilon.
\end{equation}
Here $\epsilon \in\{-1,0\}$; with $-1$ corresponding to a timelike trajectory and $0$ corresponding to a null trajectory. 
In view of the spherical symmetry we might as well just set $\theta=\pi/2$ and work with the reduced equatorial problem
\begin{eqnarray}
g_{ab} {dx^a\over d\lambda} {d x^b\over d\lambda}
&=& -e^{-2m/r} \left(dt \over d\lambda\right)^2
+ e^{2m/r} \left\{ \left(dr \over d\lambda\right)^2 + r^2\left(d\phi \over d\lambda\right)^2 \right\}
= \epsilon \in\{-1,0\}.\quad
\end{eqnarray}

The Killing symmetries imply two conserved quantities (energy and angular momentum)
\begin{equation}
e^{-2m/r} \left(dt \over d\lambda\right)=E; \qquad\qquad
e^{2m/r} r^2 \left(d\phi \over d\lambda\right)=L.
\end{equation}
Thence
\begin{equation}
e^{2m/r} \left\{-E^2 + \left(dr \over d\lambda\right)^2 \right\} + e^{-2m/r} {L^2\over r^2}
=\epsilon.
\end{equation}
That is
\begin{equation}
\left(dr \over d\lambda\right)^2 =  E^2 + e^{-2m/r} \left\{ \epsilon - e^{-2m/r} {L^2\over r^2}\right\}.
\end{equation}
This defines the ``effective potential'' for geodesic orbits
\begin{equation}
V_\epsilon(r) = e^{-2m/r} \left\{- \epsilon + e^{-2m/r} {L^2\over r^2}\right\}.
\end{equation}
\begin{itemize}
\item
For $\epsilon=0$ (massless particles such as photons), the effective potential simplifies to
\begin{equation}
V_0(r) = {e^{-4m/r} L^2\over r^2}.
\end{equation}
This has a single peak at $r=2m$ corresponding to $V_{0,max} = {L^2\over (2me)^2}$.
This is the only place where $V_0'(r)=0$, and at this point $V''(r)<0$. 
Thus there is an unstable photon sphere at $r=2m$, corresponding to the curvature coordinate $r_s = 2m \,e^{1/2} \approx 3.297442542\, m$. (This is not too far from what we would expect for Schwarzschild, where the photon sphere is at $r_s=3m$.)

For extensive discussion on the importance of the photon sphere for astrophysical imaging see for instance references~\cite{Virbhadra:1999, Virbhadra:2002, Virbhadra:1998, Virbhadra:2007,Claudel:2000, Virbhadra:2008}.

\item
For $\epsilon=-1$ (massive particles such as atoms, electrons, protons, or planets), the  effective potential is
\begin{equation}
V_1(r) =  e^{-2m/r} \left\{1 + e^{-2m/r} {L^2\over r^2}\right\}
=
e^{2W(m/r_s)} \left\{1 + {L^2\over r_s^2}\right\}.
\end{equation}
It is easy to verify that
\begin{equation}
V_1'(r) = {2 e^{-2m/r} ( L^2 e^{-2m/r} [2m-r] + m r^2) \over r^4}.
\end{equation}
and that
\begin{equation}
V_1''(r) = {2 e^{-2m/r} ( L^2 e^{-2m/r} [8m^2-12mr+3r^2] + 2m^2 r^2 -2mr^3)\over r^6}.
\end{equation}
Circular orbits, denoted $r_c$, occur at $V_1'(r)=0$, but there is no simple analytic way of determining $r_c(m,L)$ as a function of $m$ and $L$. Working more indirectly, by assuming a circular orbit ar $r=r_c$,  one can solve for the required angular momentum $L_c(r_c,m)$ as a function of $r_c$ and $m$. Explicitly:
\begin{equation}
L_c(r_c,m) =  { r_c \, e^{m/r_c} \sqrt{m} \over \sqrt{{r_c}-2m}}.
\end{equation}
(Note that at large $r_c$ we have $L_c(r_c,m)\sim \sqrt{m r_c}$ as one would expect from considering circular orbits in Newtonian gravity.)
This is enough to tell you that circular orbits for massive particles do exist all the way down to $r_c=2m$, the location of the unstable photon orbit; but this does not yet guarantee stability. 
Noting that
\begin{equation}
{\partial L_c(r_c,m)\over\partial r_c} =
{e^{m/r_c} (r_c^2 -6mr_c+4m^2) \sqrt{m} \over 2 r_c (r_c-2m)^{3/2}},
\end{equation}
we observe that the curve $L_c(r_c,m)$ has a minimum at $r_c=\left(3+\sqrt{5}\right)m$ where $L_\mathrm{min} \approx 3.523216438\, m$. (See Figure~\ref{F:angular}.) 

%---------------------------------------------------------------------------------------------
\begin{figure}[!htb]
\begin{center}
\includegraphics[scale=0.5]{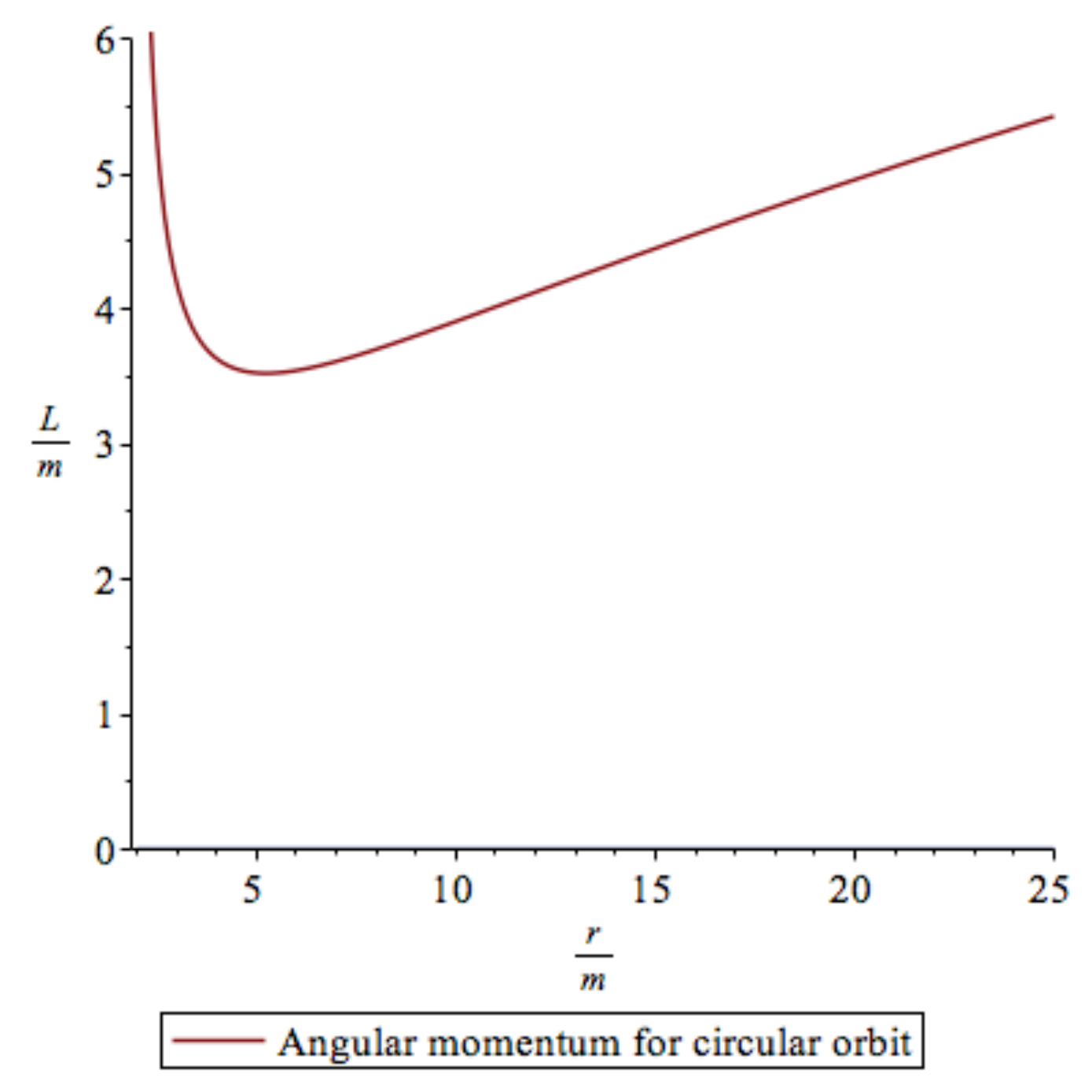}\qquad
\end{center}
{\caption{{The graph shows the angular momentum $L/m$ required to establish a circular orbit at radius $r/m$. Note the minimum at $r= (3+\sqrt{5})m$ where $L_\mathrm{min} \approx 3.523216438\, m$. Circular orbits for $r \geq (3+\sqrt{5})m$ are stable; whereas circular orbits for 
$r < (3+\sqrt{5})m$ are unstable.  (Circular orbits for $r<2m$ do not exist.)
}}\label{F:angular}}
\end{figure}
%---------------------------------------------------------------------------------------------

To check stability substitute $L_c(r_c,m)$ into $V''(r)$ to obtain
\begin{equation}
V_1''(r_c) =  {2m e^{-2m/r_c} (r_c^2 -6mr_c+4m^2) \over r_c^4 (r_c-2m)}.
\end{equation}
This changes sign when $r_c^2 -6mr_c+4m^2=0$, that is $r_c = \left(3\pm\sqrt{5}\right) m$. 
Only the positive root is relevant (the negative root lies below $r_c=2m$ where there are no circular orbits, stable or unstable).  Consequently we identify the location for the massive particle ISCO (for the exponential metric in isotropic coordinates) as
\begin{equation}
r_{\hbox{\tiny ISCO}} = \left(3+\sqrt{5}\right)m \approx 5.236067977\, m.
\end{equation}
In curvature coordinates
\begin{equation}
r_{s,\hbox{\tiny ISCO}} = \left(3+\sqrt{5}\right)\, \exp\left\{ {1\over4} \left(3-\sqrt{5}\right)\right\} \, m \approx 6.337940263\, m.
\end{equation}
This is not too far from what would have been expected in Schwarzschild spacetime, where the Schwarzschild geometry ISCO is at  $r_{s,\hbox{\tiny ISCO}}=6m$. 
\end{itemize}

%----------------------------------------------------------------------------
\section{Regge--Wheeler equation}
%----------------------------------------------------------------------------

Consider now the Regge--Wheeler equation for scalar and vector perturbations around the exponential metric spacetime. 
We will invoke the inverse Cowling approximation (wherein we keep the geometry fixed while letting the scalar and vector fields oscillate; we do this since we do not \emph{a priori} know the spacetime dynamics). The analysis closely parallels the general formalism developed in~\cite{Regge}.

Start from the exponential metric:
\begin{equation}
ds^2 = - e^{-2m/r} dt^2 + e^{+2m/r}\{dr^2 + r^2(d\theta^2+\sin^2\theta \, d\phi^2)\}.
\end{equation}
Define a tortoise coordinate by $dr_* = e^{2m/r} \; dr$ then
\begin{equation}
ds^2 =  e^{-2m/r} (-dt^2 +dr_*^2) + e^{+2m/r}r^2(d\theta^2+\sin^2\theta \, d\phi^2).
\end{equation}
Here $r$ is now implicitly a function of $r_*$.
We can also write this as 
\begin{equation}
ds^2 =  e^{-2m/r} (-dt^2 +dr_*^2) + r_s^2(d\theta^2+\sin^2\theta \, d\phi^2).
\end{equation}
Using the formalism developed in~\cite{Regge}, the Regge--Wheeler equation can be written, (using $\partial_*$ as shorthand for $\partial_{r_*}$), in the form
\begin{equation}
 \partial_*^2\, \hat \phi 
+ \left\{\omega^2- \V \right\}
  \hat \phi = 0.
\end{equation}
For a general spherically symmetric metric, (specifying the metric components in curvature coordinates), the Regge--Wheeler potential for spins $S\in\{0,1,2\}$ and angular momentum $\ell\geq S$ is~\cite{Regge}
\begin{equation}
\V_S =  (-g_{tt}) \left[{\ell(\ell+1)\over r_s^2} 
+ {S(S-1) (g^{rr}-1)\over r_s^2}\right]+ (1-S) {\partial_*^2 r_s \over r_s}.
\end{equation}
For the exponential metric in curvature coordinates we have already seen that both $g_{tt} = -e^{-2m/r}$ and $g^{rr} = (1-m/r)^2$. Therefore
\begin{equation}
\V_S =  e^{-2m/r} \left[{\ell(\ell+1)\over r_s^2} 
+ {S(S-1) [(1-m/r)^2-1]\over r_s^2}\right]+ (1-S) {\partial_*^2 r_s \over r_s}.
\end{equation}
It is important to realize that both $r_s$ and $r$ occur in the equation above. 
By noting that $\partial_* = e^{-2m/r} \partial_r$ it is possible to evaluate
\begin{equation}
 {\partial_*^2 r_s \over r_s} = {e^{-4m/r} m (2r-m)\over r^4} 
 = - {e^{-4m/r} [(1-m/r)^2-1]\over r^2},
\end{equation}
and so rephrase the Regge--Wheeler potential as
\begin{equation}
\V_S =  e^{-4m/r} \left[{\ell(\ell+1)\over r^2} 
+ {(S^2-1) [(1-m/r)^2-1]\over r^2}\right].
\end{equation}
This is always zero at $r=0$ and $r=\infty$, with some extrema at non-trivial values of $r$. 

The corresponding result for the Schwarzschild spacetime is
\begin{equation}
\V_{S,\mathrm{Sch}} =  \left(1-{2m\over r_s}\right) \left[{\ell(\ell+1)\over r_s^2} 
- {S(S-1) 2m \over r_s^3}\right]+ (1-S) {\partial_*^2 r_s \over r_s}.
\end{equation}
For the Schwarzschild metric  $\partial_* = (1-2m/r_s) \partial_{r_s}$ and so
 it is possible to evaluate
\begin{equation}
 {\partial_*^2 r_s \over r_s} =  \left(1-{2m\over r_s}\right) {2m\over r_s^3}.
\end{equation}
Then
\begin{equation}
\V_{S,\mathrm{Sch}} =  \left(1-{2m\over r_s}\right) 
\left[{\ell(\ell+1)\over r_s^2} 
- {(S^2-1) 2m \over r_s^3}\right].
\end{equation}
Converting to isotropic coordinates, which for the Schwarzschild geometry means one is applying
$r_s = r \left(1+{m\over2r}\right)^2$, we have
\begin{equation}
\V_{S,\mathrm{Sch}} =  \left(1-{m\over2r}\over1+{m\over2r}\right)^2
\left[{\ell(\ell+1)\over r^2 \left(1+{m\over2r}\right)^4 } 
- {(S^2-1) 2m \over r^3 \left(1+{m\over2r}\right)^6}\right].
\end{equation}
This is always zero at the horizon $r=m/2$ and at $r=\infty$, with some extrema at non-trivial values of $r$.

%------------------------------------------------
\subsection{Spin zero}
%------------------------------------------------
In particular for spin zero one has
\begin{eqnarray}
\V_0 
&=&  e^{-2m/r} {\,\ell(\ell+1)\over r_s^2} +  {\partial_*^2 r_s \over r_s} 
\nonumber\\
&=&  e^{-4m/r} \,{\ell(\ell+1)\over r^2} 
+  {\partial_*^2 r_s \over r_s}
\nonumber\\
&=&
e^{-4m/r} \left[{\ell(\ell+1)
- [(1-m/r)^2-1]\over r^2}\right].
\end{eqnarray}
This result can also be readily checked by brute force computation.
The corresponding result for Schwarzschild spacetime is
\begin{equation}
\V_{0,\mathrm{Sch}} =  \left(1-{m\over2r}\over1+{m\over2r}\right)^2
\left[{\ell(\ell+1)\over r^2 \left(1+{m\over2r}\right)^4 } 
+ {2m \over r^3 \left(1+{m\over2r}\right)^6}\right].
\end{equation}

\begin{figure}[!htb]
\begin{center}
\includegraphics[scale=0.5]{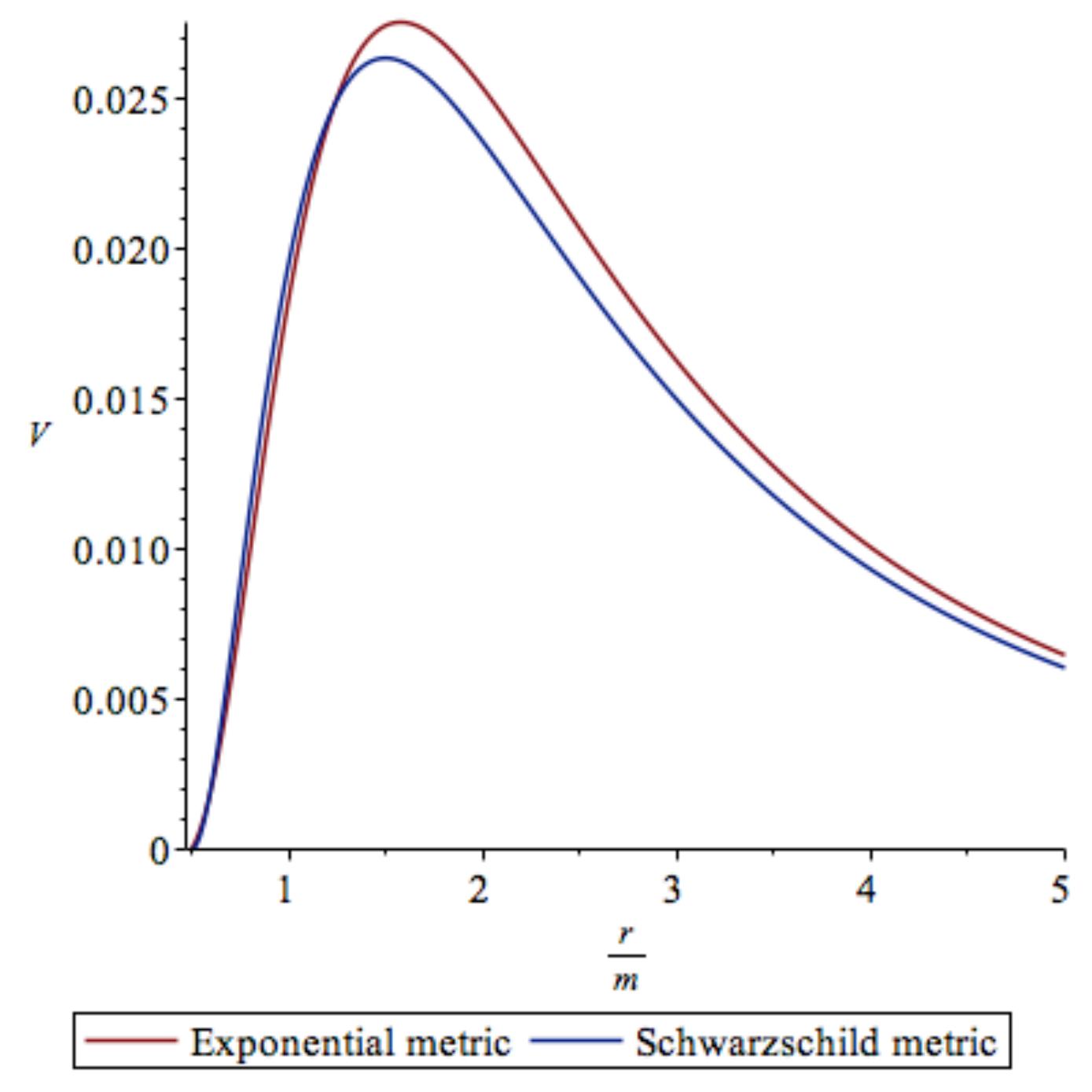}\qquad
\includegraphics[scale=0.5]{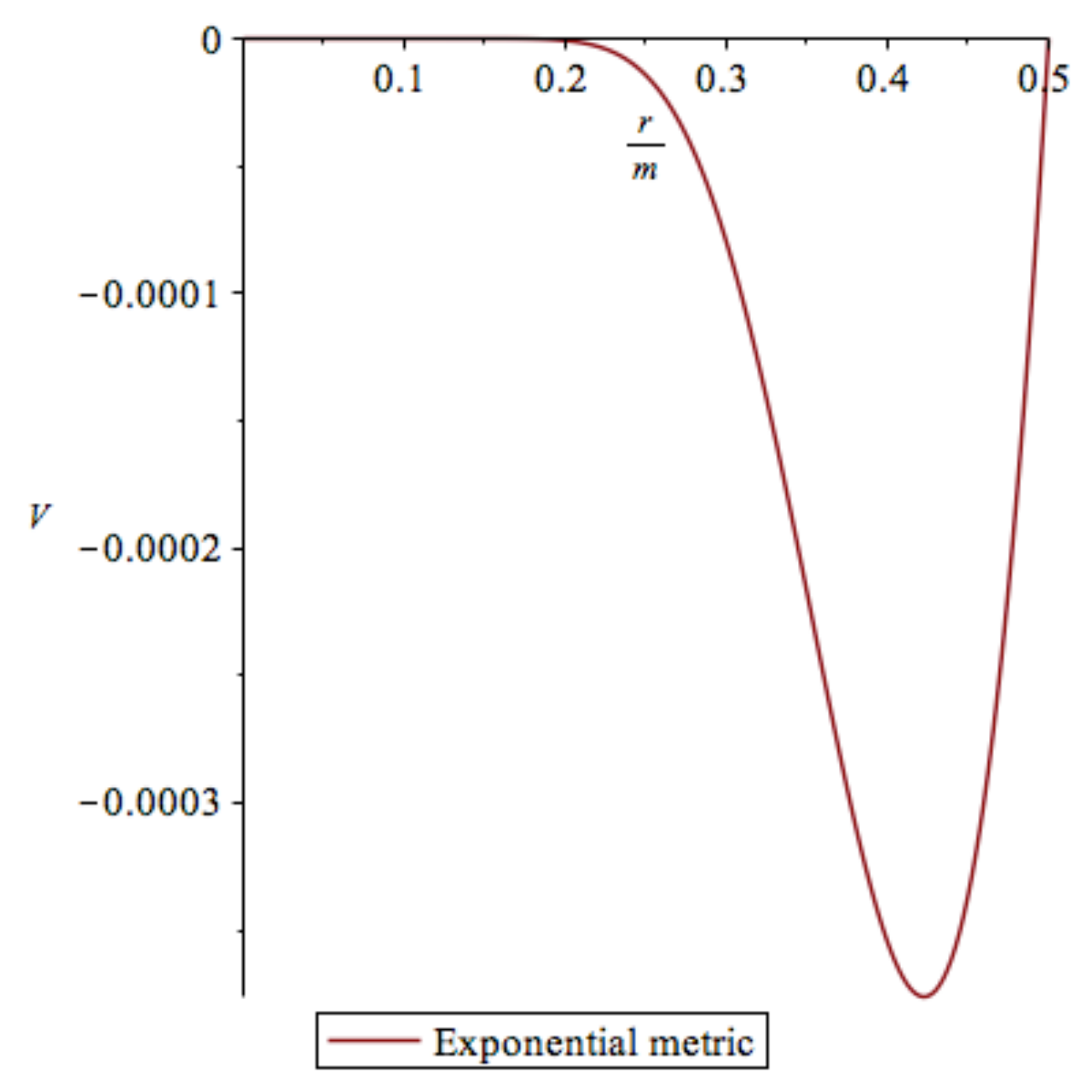}\qquad
\end{center}
{\caption{{The graph shows the spin zero Regge--Wheeler potential for $\ell=0$. 
While the Regge--Wheeler potentials are not dissimilar for $r>m/2$,
they are radically different once one goes to small $r<m/2$, (where the Regge--Wheeler potential for Schwarzschild is only formal since one is behind a horizon and cannot interact with the domain of outer communication). 
}}\label{F:V0}}
\end{figure}

For scalars the $s$-wave ($\ell=0$) is particularly important
\begin{equation}
\V_{0,\ell=0}  = e^{-4m/r} \left[ {1- (1-m/r)^2\over r^2}\right]
= e^{-4m/r} \left[{2m\over r^3 } \left(1-{m\over2r} \right)\right];
\end{equation}
versus
\begin{equation}
\V_{0,\ell=0,\mathrm{Sch}} =\left(1-{m\over2r}\over1+{m\over2r}\right)^2
\left[ {2m \over r^3 \left(1+{m\over2r}\right)^6}\right].
\end{equation}
Note that these potentials both have zeros at $r=m/2$ and that for $r<m/2$ only the exponential Regge--Wheeler potential is of physical interest, (thanks to the horizon at $r=m/2$ in the Schwarzschild metric). See Figure~\ref{F:V0}. The potential peaks are at $r=\left(1+{1\over\sqrt{3}}\right)m$ and $r={3m\over2}$ respectively.  For the exponential metric there is also a trough (a local minimum) at $r=\left(1-{1\over\sqrt{3}}\right)m$. 

%----------------------------------------------------------------------------
\subsection{Spin one}
%----------------------------------------------------------------------------

\begin{figure}[!htb]
\begin{center}
\includegraphics[scale=0.5]{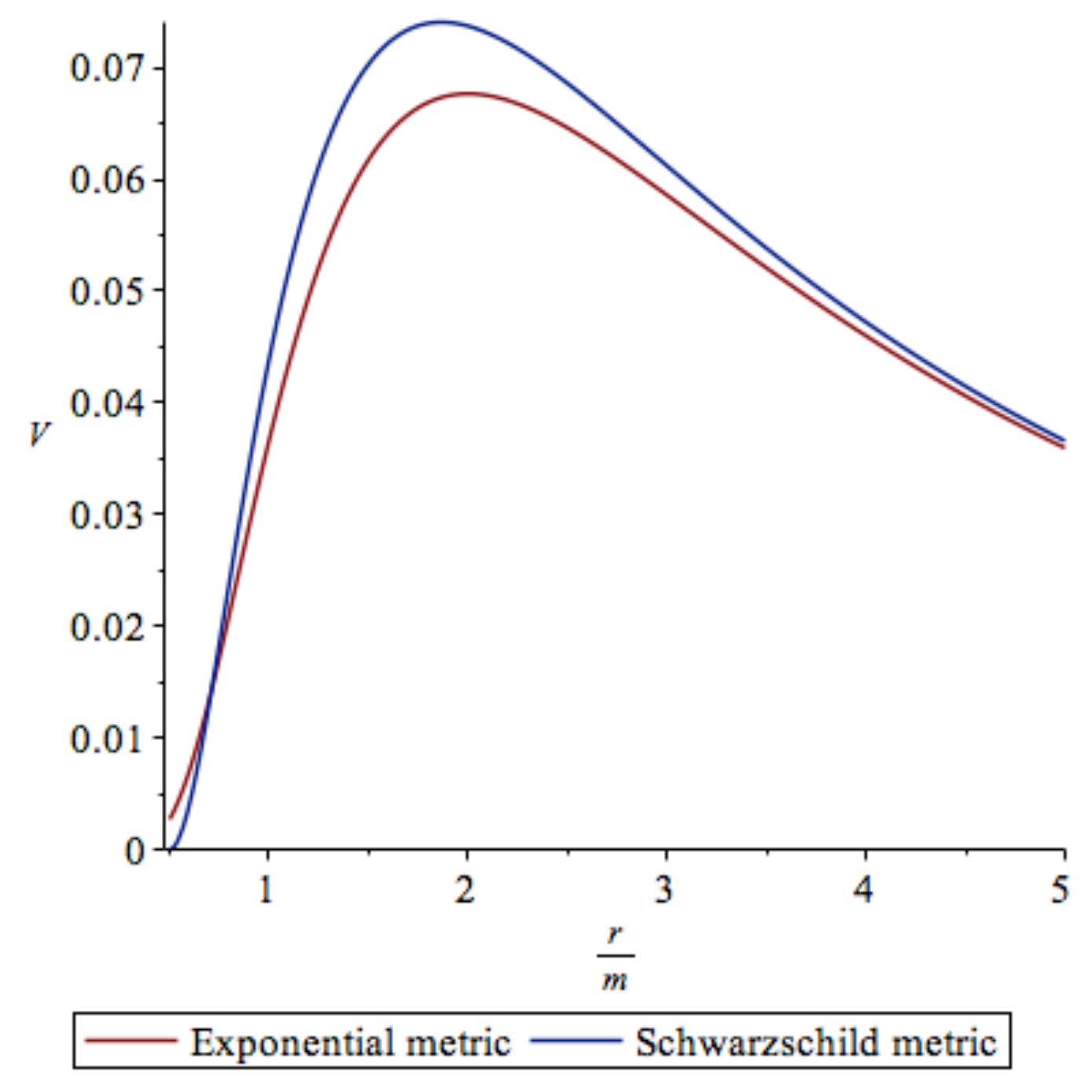}\qquad
\includegraphics[scale=0.5]{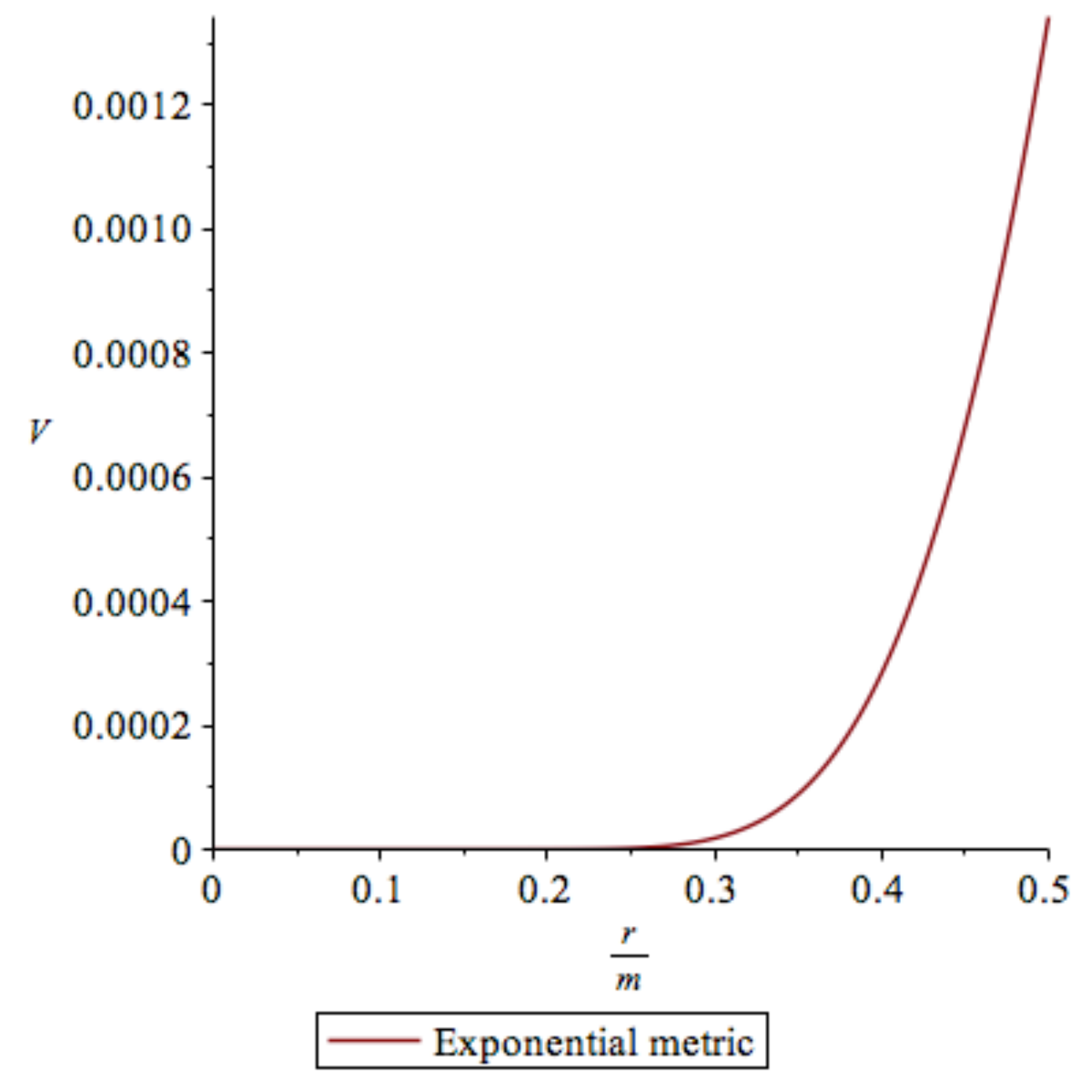}\qquad
\end{center}
{\caption{{The graph shows the spin one Regge--Wheeler potential for $\ell=1$. 
While the Regge--Wheeler potentials are not dissimilar for $r>m/2$,
they are radically different once one goes to small $r<m/2$, (where the Regge--Wheeler potential for Schwarzschild is only formal since one is behind a horizon and cannot interact with the domain of outer communication). 
}}\label{F:V1}}
\end{figure}

For the spin one vector field the $\left\{ {r_s^{-1}} \partial_{*}^2{ r_s} \right\} $ term drops out; this can ultimately be traced back to the conformal invariance of massless spin 1 particles in (3+1) dimensions. We are left with the particularly simple result ($\ell\geq 1$)
\begin{equation}
\V_1 =  {e^{-4m/r} \ell(\ell+1)\over r^2}.
\end{equation}
See related brief comments regarding conformal invariance in~\cite{Regge}.
Note that this rises from zero ($r\to0$) to some maximum at $r=2m$, where $\V_1\to {\ell(\ell+1)\over(2me)^2}$ and then decays back to zero ($r\to\infty$).
The corresponding result for Schwarzschild spacetime is
\begin{equation}
\V_{1,\mathrm{Sch}} =  {\left(1-{m\over2r}\right)^2\over\left(1+{m\over2r}\right)^6} \; 
{\ell(\ell+1)\over r^2}.
\end{equation}
Note that this rises from zero (at $r= m/2$) to some maximum at $r=\left(1+{\sqrt{3}\over2}\right) m$, where $\V_1\to {2\ell(\ell+1)\over27m^2}$, and then decays back to zero ($r\to\infty$).
See Figure \ref{F:V1} for qualitative features of the potential.

%---------------------------------------------------------
%-----------------------------------------------------------

%----------------------------------------------------------------------------
\subsection{Spin two}
%----------------------------------------------------------------------------

\begin{figure}[!htb]
\begin{center}
\includegraphics[scale=0.5]{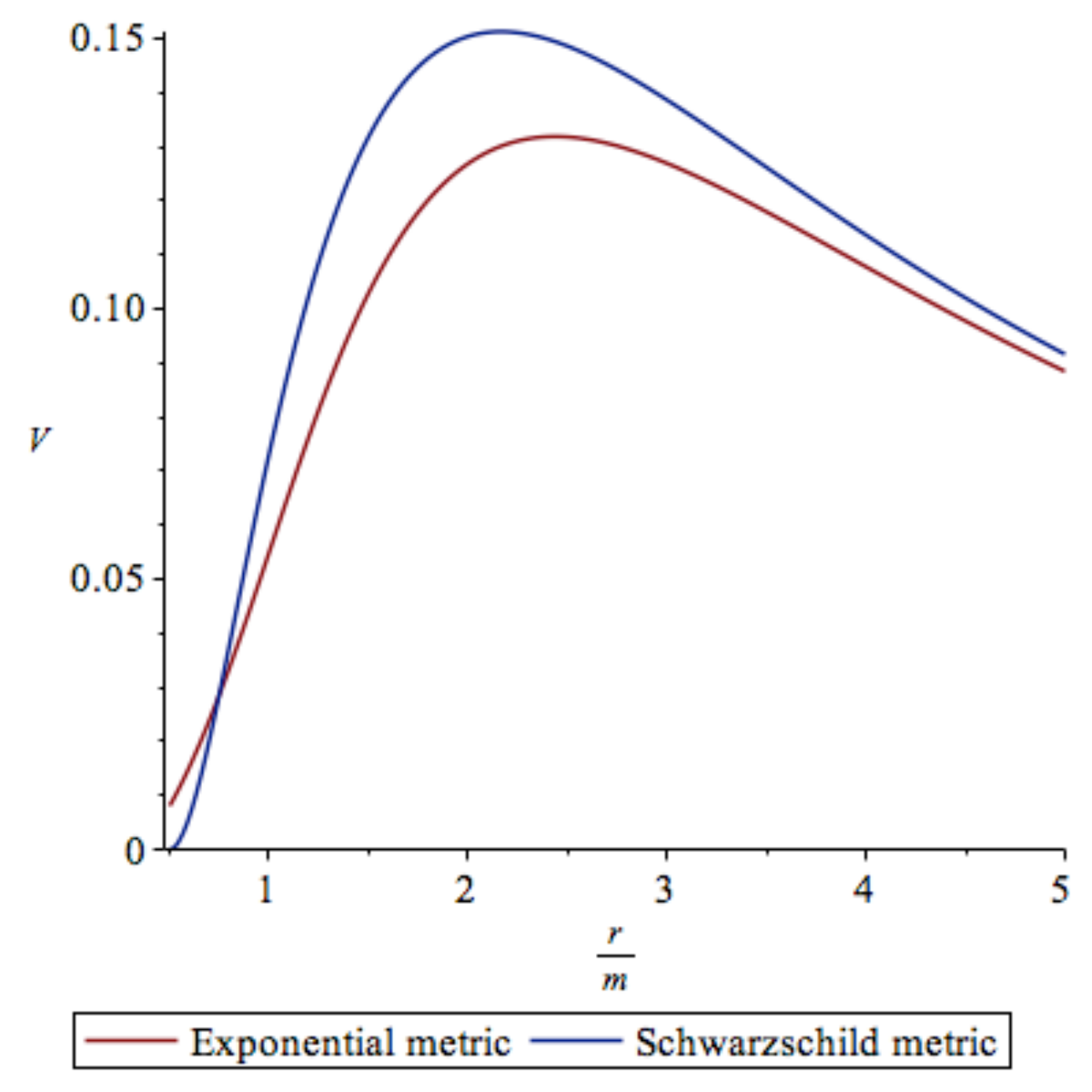}\qquad
\includegraphics[scale=0.5]{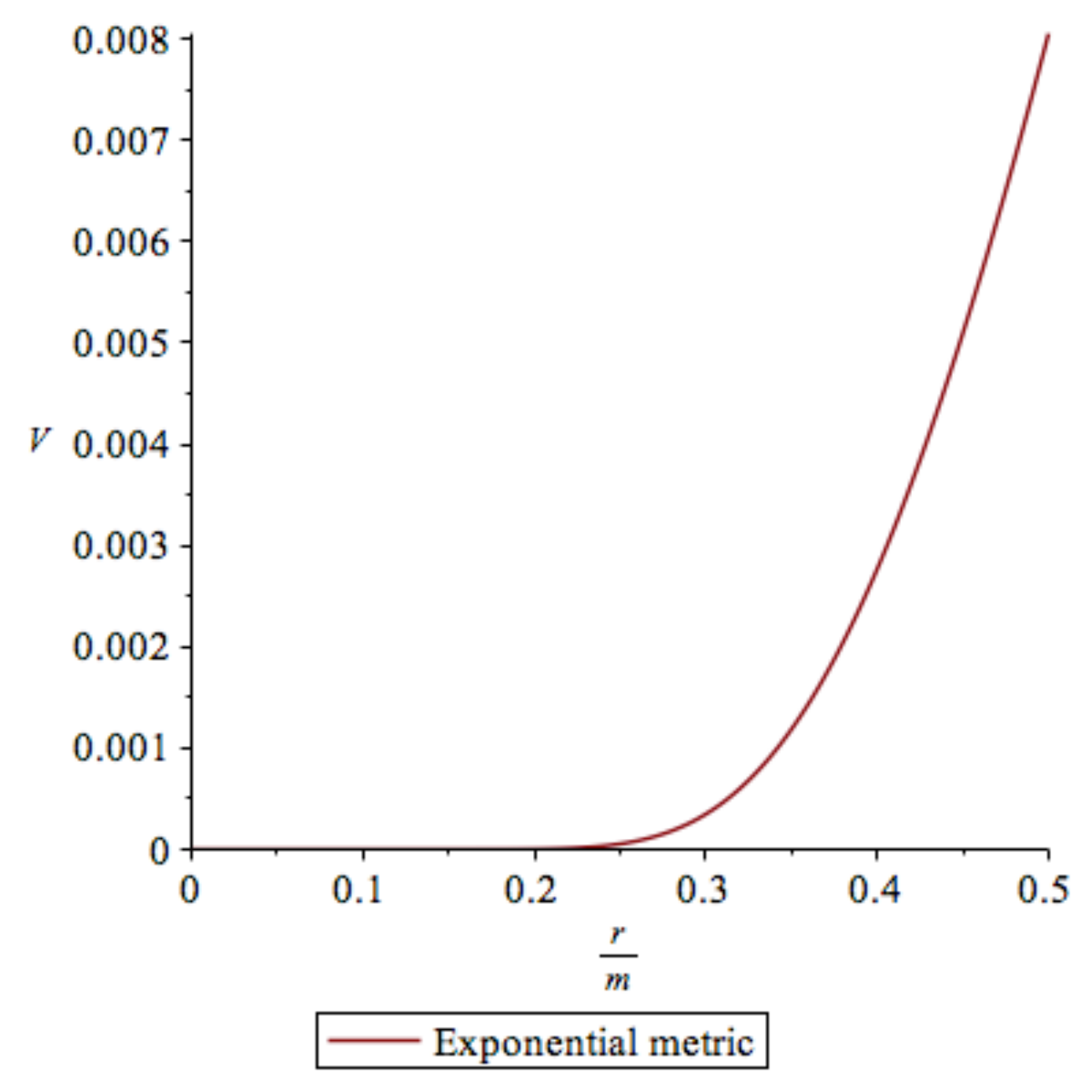}\qquad
\end{center}
{\caption{{The graph shows the spin two (axial) Regge--Wheeler potential for $\ell=2$. 
The Regge--Wheeler potentials are somewhat dissimilar for $r>m/2$,
and are radically different once one goes to small $r<m/2$, (where the Regge--Wheeler potential for Schwarzschild is only formal since one is behind a horizon and cannot interact with the domain of outer communication). 
}}\label{F:V2}}
\end{figure}

For spin two, more precisely for spin 2 axial perturbations, (see~\cite{Regge}) one has ($\ell\geq 2$)
\begin{eqnarray}
\V_2 
&=&  e^{-2m/r} {\,\ell(\ell+1)\over r_s^2} -3  \,{\partial_*^2 r_s \over r_s} 
\nonumber\\
&=&  e^{-4m/r} \,{\ell(\ell+1)\over r^2} 
-3 \, {\partial_*^2 r_s \over r_s}
\nonumber\\
&=&
e^{-4m/r} \left[{\ell(\ell+1)
+3 [(1-m/r)^2-1]\over r^2}\right].
\end{eqnarray}
The corresponding result for Schwarzschild spacetime is
\begin{equation}
\V_{2,\mathrm{Sch}} =  \left(1-{m\over2r}\over1+{m\over2r}\right)^2
\left[{\ell(\ell+1)\over r^2 \left(1+{m\over2r}\right)^4 } 
- {6m \over r^3 \left(1+{m\over2r}\right)^6}\right].
\end{equation}
See Figure \ref{F:V2} for qualitative features of the potential.

%-----------------------------------------------------------------
%-------------------------------------------------------------------

\vspace{2cm}
%------------------------------------------------
\section{GR interpretation for the exponential metric}\label{S:GR}
%------------------------------------------------

\enlargethispage{10pt}
While many of the proponents of the exponential metric have for one reason or another been trying to build ``alternatives'' to standard general relativity, there is nevertheless a relatively simple interpretation of the exponential metric within the framework of standard general relativity and the standard Einstein equations, albeit with an ``exotic'' matter source. The key starting point is to note:
\begin{equation}
R_{ab} 
= -{2m^2 \over r^4} \; \text{diag}\{0,1,0,0\}_{ab} 
= - {1\over2} \nabla_a\left(2m\over r\right) \nabla_b\left(2m\over r\right) 
= - {1\over2} \nabla_a \Phi \, \nabla_b \Phi.
\end{equation}
Equivalently
\begin{equation}
G_{ab} 
= - {1\over2} \left\{ \nabla_a \Phi \, \nabla_b \Phi - {1\over 2}  g_{ab} \, (g^{cd} \nabla_c\Phi \nabla_d \Phi) \right\}.
\end{equation}
This is just the usual Einstein equation for a \emph{negative kinetic energy massless scalar field}, a ``ghost'' or ``phantom'' field. The contracted Bianchi identity $G^{ab}{}_{;b}$ then automatically yields the scalar field EOM $(g^{ab} \nabla_a\nabla_b) \Phi=0$. That the scalar field has negative kinetic energy is intimately related to the fact that the exponential metric describes a traversable wormhole~\cite{Morris-Thorne,Visser:book}.

So, perhaps ironically, despite the fact that many of the proponents of the exponential metric for one reason or another reject general relativity, the exponential metric they advocate has a straightforward if somewhat exotic general relativistic interpretation.\footnote{It is also possible to interpret the exponential metric as a special sub-case of the Brans class IV solution of Brans--Dicke theory, which in turn is a special case of the general spherical, asymptotically flat, vacuum solution~\cite{Faraoni:2016,Faraoni:2018}; in this context it is indeed known that some solutions admit a wormhole throat, but that message seems not to have reached the wider community.
}

%------------------------------------------------
\section{Discussion}\label{S:Discussion}
%------------------------------------------------

Regardless of one's views regarding the merits of some of the ``justifications'' used for advocating the exponential metric, the exponential metric can simply be viewed as a phenomenological model that can be studied in its own right.  Viewed in this way the exponential metric has a number of interesting features:
\begin{itemize}
\item It is a traversable wormhole, with time slowed down on the other side of the wormhole throat. 
\item Strong field lensing phenomena are markedly different from Schwarzschild.
\item ISCOs and unstable photon orbits still exist, and are moderately shifted from where they would be located in Schwarzschild spacetime.
\item  Regge--Wheeler potentials can still be extracted,  and are moderately different from what they would be in Schwarzschild spacetime.
\end{itemize}
Many of the proponents of the exponential metric are arguing for using it as a replacement for the Schwarzschild geometry of general relativity --- however typically without any detailed assessment of the phenomenology.
We strongly feel that if one wishes to replace all the black hole candidates astronomers have identified with traversable wormholes, then certainly a careful phenomenological analysis of this quite radical proposal (somewhat along the lines above) should be carried out.
Perhaps most ironically, despite the fact that many of the proponents of the exponential metric reject general relativity, the exponential metric has a natural interpretation in terms of general relativity coupled to a phantom scalar field.

\clearpage
%------------------------------------------------
\section*{Acknowledgements}
%----------------------------------------------------------------------------
This project was funded by the Ratchadapisek Sompoch Endowment Fund, Chulalongkorn University (Sci-Super 2014-032), by a grant for the professional development of new academic staff from the Ratchadapisek Somphot Fund at Chulalongkorn University, by the Thailand Research Fund (TRF), and by the Office of the Higher Education Commission (OHEC), Faculty of Science, Chulalongkorn University (RSA5980038). PB was additionally supported by a scholarship from the Royal Government of Thailand. TN was also additionally supported by a scholarship from the Development and Promotion of Science and Technology talent project (DPST). MV was supported by the Marsden Fund, via a grant administered by the Royal Society of New Zealand.

The authors wish to thank Kumar Virbhadra and Valerio Faraoni for their interest and comments. 

%----------------------------------------------------------------------------

%----------------------------------------------------------------------------
\end{document}